# Intrinsic mechanism for high temperature ferromagnetism in GaMnN

Saki Sonoda*[1], Isao Tanaka[2], Hidekazu Ikeno[2], Tomoyuki Yamamoto[3], Fumiyasu Oba[2], Tsutomu Araki[4], Yoshiyuki Yamamoto[5], Ken-ichi Suga[6], Yasushi Nanishi[4], Youichi Akasaka[1], Koichi Kindo[6,7], & Hidenobu Hori[5]

[1]*Department of System Innovation, Osaka University,1-3 Machikaneyama, Toyonaka, Osaka 560-8531 Japan*, [2]*Department of Material Science and Engineering, Kyoto University, Yoshidahonmachi, Sakyo, Kyoto 606-8501 Japan*, [3]*Department of Material Science and Engineering, Waseda University, 3-4-1 Okubo, Shinjuku, Tokyo 169-8555 Japan*, [4]*Department of Photonics, Ritsumeikan University,1-1-1 Norohigashi, Kusatsu, Shiga 525-8577 Japan*, [5]*School of Material Science, Japan Advanced Institute of Science and Technology, 1-1 Asahidai, Tatsunokuchi, Nomi, Ishikawa 923-1292 Japan*, [6]*KYOKUGEN, Osaka University,1-3 Machikaneyama, Toyonaka, Osaka 560-8531 Japan*, [7]*The Institute for Solid State Physics, University of Tokyo, 5-1-5 Kashiwanoha Kashiwa, Chiba 277-8581 Japan*

* fax: +81-6-6850-6341     e-mail: saki@nano.ee.es.osaka-u.ac.jp.

Considerable efforts have been devoted recently to synthesize diluted magnetic semiconductors having ferromagnetic properties at room temperature because of their technological impacts for spintronic devices. In 2001 successful growth of GaMnN films showing room temperature ferromagnetism and p-type conductivity has been reported.[1,2]. The estimated Curie temperature was 940K at 5.7% of Mn, which is highest among diluted magnetic semiconductors ever been reported. However, the electronic mechanism behind the ferromagnetic behaviour has still been controversial. Here we show experimental evidence using the ferromagnetic samples that Mn atoms are substitutionally dissolved into the GaN lattice and they exhibit mixed valences of +2 (majority) and +3 (minority). The p-type carrier density decreases significantly at very low temperatures. At the same time, magnetization dramatically decreases. The results imply that the ferromagnetic coupling between Mn atoms is mediated by holes in the mid-gap Mn band.

Fundamental knowledge on the electronic structure of transition metal elements in host semiconductors is essential for design and development of new spintronic devices. Ferromagnetic III-V diluted magnetic semiconductor (DMS) was first recognized in InMnAs and GaMnAs.[3-5] Mean-field Zener model has been adopted to explain the ferromagnetism in these systems.[6] Based on the model, room temperature ferromagnetism in GaMnN and ZnMnO has been predicted.[6] After the discovery of GaMnN with an extremely high Curie temperature, $T_C$,[1,7] the mechanism of ferromagnetism in GaMnN has been discussed from comparative view with GaMnAs.[8-10] Dietl *et al.*[11] proposed that the high temperature ferromagnetism in GaMnN can be explained within their mean-field Zener model, similar to the case of GaMnAs. This model is based on the idea that the $Mn^{3+}/Mn^{2+}$ level is located within or near the valence band and holes are thereby introduced. Mn should be in the charge state of $Mn^{2+}$. The experimental result by Mn-K NEXAFS (near edge x-ray absorption fine structures)[12] was taken as an evidence for $Mn^{2+}$ in GaMnN. Mn-$L_{2,3}$ NEXAFS of GaMnN[13] seems to support the presence of $Mn^{2+}$. On the other hand, first principles calculations have predicted that the local electronic structure of Mn in GaN significantly differs from Mn in GaAs.[10, 14, 15] Deep $Mn^{3+}/Mn^{2+}$ level appears in the band gap in GaN, while the $Mn^{3+}/Mn^{2+}$ state in GaAs is shallow making significant overlap between Mn-3d and As-4p valence states. However, a recent first principles study failed to predict high temperature ferromagnetism in GaMnN.[15]

A typical argument for the unexplained ferromagnetism has been formation of ferromagnetic clusters or segregations. Rao and Jena[16] proposed that the formation of

$Mn_xN$ type clusters brings about the ferromagnetism through first principles calculations. More recent calculations, however, suggested that such clustering decreases the $T_C$[18]. It is true that Pearton et al.[17] reported that $Ga_xMn_y$ type crystalline secondary phases having $T_C$ exceeding room temperature were found by x-ray diffraction (XRD) in GaMnN film (5%Mn) when grown under "unoptimized" conditions. However, the formation of such clusters has never been verified by XRD in our ferromagnetic samples. The presence of nano-scale ferromagnetic inclusions cannot be ruled out only by the XRD result. In the present study, we adopt x-ray absorption spectroscopy to identify the physical and chemical states of Mn atoms.

The GaMnN films in the present study were grown by molecular beam epitaxy using $NH_3$ as nitrogen source. The procedure is the same as those reported in earlier papers.[1,2] Here we report results for thicker films as 1μm and higher Mn concentrations up to 8.2 at% as determined by electron probe microanalyzer (EPMA). XRD shown in Fig.1a does not exhibit any crystalline secondary phases even in the 8.2at%Mn-film.

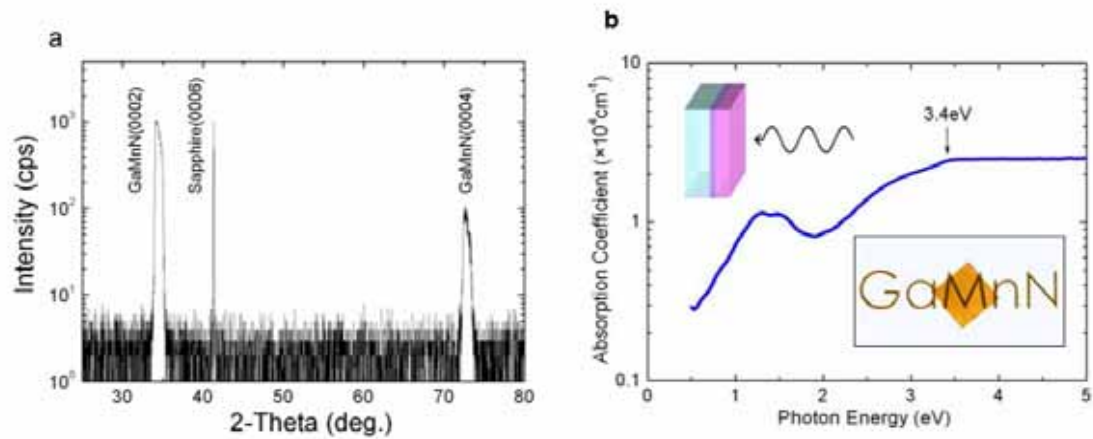

Figure 1 **a**, XRD patterns of the GaMnN film (Mn:8.2%) (θ-2θ scan) and **b**, Optical absorption spectrum of the GaMnN (Mn:8.2%) with a photograph of the present GaMnN film. Although the film is as thick as 1μm, letters below the film can be read because of its good transparency.

The yellowish film of 1μm in thickness remains transparent as can be seen in the photo in Fig. 1b. Optical absorption spectra obtained at room temperature is also shown in Fig. 1b. Corresponding to its yellowish colour, an absorption peak was observed at around 1.5 eV followed by a broad absorption started at around 1.8eV. The spectrum is similar to those reported by Graf *et al.* [23] on a GaN:Mn film with 0.23% Mn, and that by Wolos *et al.*[24] on a GaN:Mn+Mg bulk crystal with 0.009% Mn. The peak at around 1.5eV was found to be absent in Si-codoped GaN:Mn in which Si atoms were expected to act as electron donors.[23] The 1.5eV-peak is a signature of the presence of $Mn^{3+}$, since $Mn^{2+}$ with a half filled 3d levels does not contribute for spin-allowed d-d absorption in this energy range.

NEXAFS provides element-specific information of local environment of atoms. Samples for NEXAFS need not be bulk crystals. Nano-crystalline or amorphous materials can also be examined. In the present study, we made a series of NEXAFS experiments on our ferromagnetic samples and analysed the results with the aid of first principles calculations. $Mn-L_{2,3}$ NEXAFS of the films were measured at ALS (Advanced light source, Berkeley) BL8.0.1 with the total electron yield method at room temperature. The shape of the spectrum is found to be independent of the Mn concentration. A typical spectrum shown in Fig. 2a is the same as that reported by Edmonds *et al.* [13] for GaMnN with 3%Mn. However, no detailed information on the properties of their sample was available. No interpretation of the spectrum was provided, either, except for that the valence state of Mn is +2, based on finger printings of some Mn compounds. In the present study, we use a novel first principles multi-electron method to calculate the $Mn-L_{2,3}$ NEXAFS. Because of the strong correlation among 3d electrons and a 2p core-hole, the spectrum shows clear multiplet structures that cannot be reproduced by a simple one-electron calculation. In the present study, we firstly optimized atomic positions by a plane-wave pseudopotentials method[19] using supercells of $(Ga_{35}Mn_1N_{36})$. Then the multi-electron calculations have been made employing all configuration interactions among Mn-3d and N-2p electrons together with Mn-2p core electrons using $MnN_4^{9-}$ and $MnN_4^{10-}$ models that correspond to $Mn^{3+}$ and $Mn^{2+}$, respectively. Details of the multi-electron method can be found elsewhere.[20] Comparison of theoretical spectra with experimental one finds that the major part of Mn atoms is substitutionally dissolved into GaN as $Mn^{2+}$. Similar comparison has been made for Mn-$L_{2,3}$ NEXAFS of ZnMnO with 5%Mn that shows no ferromagnetism at room temperature. Excellent agreement between experimental and theoretical spectra with $Mn^{2+}$ can be seen in ZnMnO. Small but clear difference between experimental spectra of ZnMnO and GaMnN is recognized at the small satellite peak as indicated by arrows in Fig. 2a. This can be ascribed to the presence of $Mn^{3+}$ component only in GaMnN. Indeed, the experimental spectrum of GaMnN is better reproduced when 15% of Mn atoms are assumed to be in the 3+ state. The difference between magnetic properties of ZnMnO and GaMnN can be ascribed to the presence of the $Mn^{3+}$ component. On the basis of two spectroscopic results, we can conclude that $Mn^{2+}$ and $Mn^{3+}$ are coexistent in the present GaMnN sample with $Mn^{2+}$ as a major component.

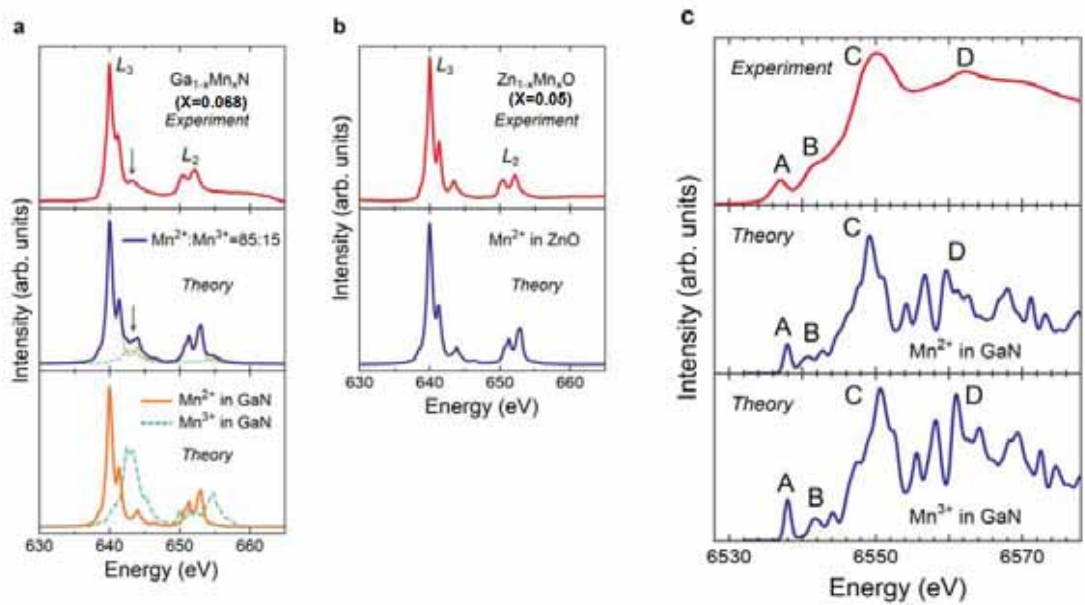

Figure 2 **a**, Mn $L_{2,3}$-NEXAFS of GaMnN (Mn:6.8%). Top panel shows an experimental spectrum. Bottom one presents calculated spectra for $Mn^{2+}$ (orange) and $Mn^{3+}$ (green) using the first principles multi-electron method. Middle one is composed of 15% of $Mn^{3+}$ and 85% of $Mn^{2+}$ calculated spectra, which shows the best agreement to the experimental spectrum. **b**, Mn $L_{2,3}$-NEXAFS of ZnMnO with 5%Mn. **c**, Mn K-NEXAFS of GaMnN. Lower two panels are calculated spectrum for $Mn^{2+}$ and $Mn^{3+}$ in GaN using FLAPW+lo method [21] and upper panel is experimental spectrum.

Mn-K NEXAFS of the film was also measured and compared with theoretical calculations. (Fig.2c) [21] It well agrees with those in literature.[12, 22] However, clear interpretation of the experimental spectrum has not been provided thus far. Main conclusions by the Mn-K NEXAFS are two fold: (1) excellent agreement between theory and experiment shows that Mn atoms are substitutionally present at Ga sites. Signature of other sites of Mn, such as in metallic or clustered Mn, is not evident. Soo *et al*. [12] mentioned that the peak A may be ascribed to Mn clusters. This assertion can be discounted because peak A is clearly seen in the present theoretical spectrum. (2) Although Mn at different hosts shows different fingerprints in general, it is very difficult to distinguish the difference in valence between $Mn^{2+}$ and $Mn^{3+}$ in GaMnN only by the Mn-K NEXAFS if they are present at the substitutional sites. Instead, Mn-$L_{2,3}$ NEXAFS can reveal the valence states with the assistance of reliable theoretical calculations.

Both of results by optical absorption and $L_{2,3}$ NEXAFS unambiguously show the coexistent of the two valence states in the present GaMnN samples. This implies that the electronic level of $Mn^{2+}/Mn^{3+}$ stays in the mid-gap, which is consistent to the results by first principles calculations.[14] The assumption of the mean field Zener model is therefore not valid in the case of GaMnN.

Figure 3a shows the magnetic field dependence of magnetization (M-H curve) of the 6.8at%Mn sample measured by superconducting quantum interference device (SQUID) magnetometer at 400K in the magnetic field parallel and perpendicular to the film plane. As can be seen, the film is ferromagnetic even at 400K with the easy axis in the plane.

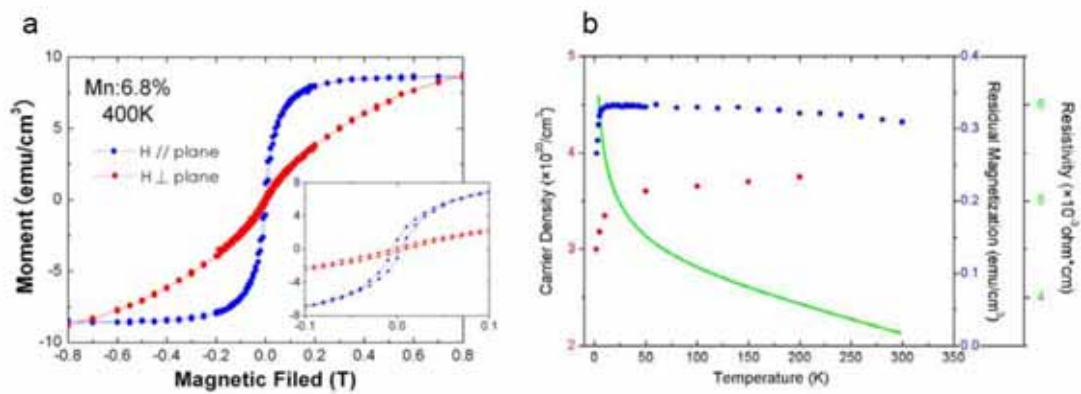

Figure3 **a**, Magnetic field dependence of magnetization (M-H curve) of GaMnN (Mn:6.8%) film at 400K in the magnetic field parallel and perpendicular to the film. The inset is the magnified M-H curve at around 0T. **b**, Temperature dependences of residual magnetization (M-T curves) of the GaMnN film (Mn:6.8%), and its carrier density and electric resistivity.

In order to find out the origin of ferromagnetism, temperature dependence of the magnetization has been examined. Firstly, magnetic field of 1T was applied parallel to the film at 300K. At this temperature, this film is fully magnetized with 1T. After decreasing the applied magnetic field down to 0T, magnetization of this film was measured while cooling it without the external magnetic field (blue dots in Fig. 3b). Down to around 10K, the magnetization remains almost unchanged. But it drops significantly with further decrease of temperature. The magnetization behaviour is almost reversible against temperature. Such a temperature dependence of the residual magnetization in a GaMnN sample is reported for the first time in this study. In parallel to the magnetic measurements, electric transport has been examined. The resistance was measured by the four-terminal method, and the carrier density was evaluated from Hall resistance. The major carrier in these samples is p-type. The carrier concentration (red dots in Fig.3b) is found to decrease remarkably below 10K. The electric resistively of the sample as shown by a green curve shows the reciprocal tendency.

The decrease of the residual magnetization in the low temperature region corresponds well to the decrease of the hole density, which clearly implies good correlation between ferromagnetism and mobile holes. It should be emphasized that the weakening of the ferromagnetism at the low temperature and the reversibility of magnetization with temperature cannot be explained by the presence of ferromagnetic clusters such as $Mn_xN_y$ or $Ga_xMn_y$.

Because $Mn^{2+}$ and $Mn^{3+}$ are coexsistent, Fermi level should be located at the Mn-3d band as in the case of colossal magnetoresistance manganites, not in the valence band. The p-type conductivity in the present sample may take place with d-electrons hopping between $Mn^{2+}$ and $Mn^{3+}$. The ferromagnetic coupling between Mn atoms may be driven by a double exchange that is mediated by the hole in the Mn band. The carrier density of $3.8 \times 10^{20}$ cm$^{-3}$ corresponds to about 13% of total Mn atoms in the 6.8at%Mn sample. This agrees well with the fraction of $Mn^{3+}$ as assumed for the analysis of the Mn-$L_{2,3}$ NEXAFS. The saturation magnetization is $0.3\mu_B$/Mn when the magnetic field is applied parallel to the film. This corresponds to $2.1\mu_B$ per a hole or an $Mn^{3+}$ ion.

The reader may be puzzled by the coexistence of $Mn^{2+}$ and $Mn^{3+}$ in p-type GaMnN materials since the $Mn^{2+}/Mn^{3+}$ level is deep from the top of the valence band. It is true that Mn should be in the charge of +3 not +2 merely within this scenario. We have to think of the presence of donor-type defects that contribute to the formation of $Mn^{2+}$. There are at least two strong candidates for such donors. One is the nitrogen vacancy, $V_N$. Recent careful first principles calculations have concluded that positive charge states are energetically favourable for $V_N$ irrespective of the Fermi level, indicating that it always acts as a donor or compensates acceptor-type defects.[25] A high-energy electron irradiation study of hydride vapour phase epitaxy-grown GaN films has also shown the donor nature of $V_N$.[26] Another candidate is hydrogen. As in the case of $V_N$, donor-like behaviour of hydrogen has been suggested by both experimental and theoretical works.[27,28] A large amount of hydrogen is known to be incorporated in GaN films grown by MOCVD utilizing $NH_3$ as nitrogen source and $H_2$ as carrier gas. In fact, we have observed existence of $10^{19}/cm^3$ of hydrogen in the film with 8.2% of Mn by SIMS (secondary ion mass spectroscopy). Baik et al.[29] examined the effect of H in GaMnN on the magnetization. They irradiated H plasma to n-type GaMnN film with 3% Mn and indeed observed the increase of residual magnetization. This behaviour seems to support the mechanism that was recently proposed by Coey et al.[30] who assumed the presence of long range ferromagnetic coupling via donors associated with the n-type conductivity. However, their theory cannot be used to explain our experimental data, because our samples show clear p-type conductivity rather than n-type.

The present study has confirmed that the high temperature ferromagnetism of GaMnN is an intrinsic behaviour to Mn ions substitutionally dissolved into GaN. There is no need to think of the external source of ferromagnetism such as precipitates. Coexistence of $Mn^{2+}/Mn^{3+}$ resulting in the p-type conductivity via d-electrons hopping is suggested to play very important role for the ferromagnetism. The mechanism we propose may be valid for other wide-gap semiconductors doped with transition elements showing mid-gap redox levels and p-type conductivity thereon.

REFERENCES

[1] S. Sonoda, S. Shimizu, T. Sasaki, Y. Yamamoto, H. Hori, *J. Cryst. Growt.* **2002**, *237-239*, 1358.

[2] H. Hori, S. Sonoda, T. Sasaki, Y. Yamamoto, S. Shimizu, K. Suga, K. Kindo, *Physica B* **2002**, *324*, 142.

[3] H. Munekata, H. Ohno, S. von Molnar, A. Segmüller, L. L. Chang, L. Esaki, *Phys. Rev. Lett.* **1989**, *63*, 1849.

[4] H. Ohno, H. Munekata, T. Penney, S. von Molnar, L. L. Chang, *Phys. Rev. Lett.* **1992**, *68*, 2664.

[5] H. Ohno, A. Shen, F. Matsukura A. Oiwa, A. Endo, S. Katsumoto, Y. Iye, *Appl. Phys. Lett*. **1996**, *69*, 363.

[6] T. Dietl, H. Ohno, F. Matsukura, J. Cibert, D. Ferrand, *Science* **2000**, *287*, 1019.

[7] M. L. Reed, *et a*l. M. L. Reed, N. A. El-Masry, H. H. Stadelmaier, M. K. Ritums, M. J. Reed, C. A. Parker, J. C. Roberts, S. M. Bedair, *Appl. Phys. Lett*. **2001**, *79*, 3473.

[8] L. Kronik, M. Jain, J. R. Chelikowsky, *Phys. Rev. B* **2002**, *66*, 041203.

[9] E. Kulatov, H. Nakayama, H. Mariette, H. Ohta, Y. A. Uspenskii, *Phys. Rev. B* **2002**, *66*, 045203.

[10] B. Sanyal, O. Benqone, S. Mirbt, *Phys. Rev. B* **2003**, *68*, 205210.

[11] T. Dietl, F. Matsukura, H. Ohno, *Phys. Rev. B* **2002**, *66*, 033203.

[12] Y. L. Soo, G. Kioseoglou, S. Kim, S. Huang, and Y. H. Kao, S. Kuwabara, S. Owa, T. Kondo, H. Munekata, *Appl. Phys. Lett*. **2001**, *79*, 3926.

[13] K. W. Edmonds, *J. Appl. Phys*. **2004**, *95*, 7166.

[14] K. Sato, H. Katayama-Yoshida, *Semicond. Sci. Technol*. **2002**, *17*, 367.

[15] K. Sato, W. Schweika and P. H. Dederichs, H. Katayama-Yoshida *Phys. Rev. B* **2004**, *70*, 201202R.

[16] B. K. Rao, P. Jena, *Phys. Rev. Lett*. **2002**, *89*, 185504.

[17] S. J. Pearton, C. R. Abernathy, G. T. Thaler, R. M. Frazier, D. P. Norton, F. Ren, Y. D. Park, J. M. Zavada, I. A. Buyanova, W. M. Chen A. F. Hebard, *J. Phys. Condens. Matter* **2004**, *16*, R209.

[18] L. M. Sandratskii, P. Bruno, S. Mirbt, *Phys. Rev. B* **2005**, *71*, 045210.

[19] V. Milman, B. Winkler, J. A. White, C. J. Pickard, M. C. Payne, E. V. Akhmatskaya, R. H. Nobes, *Int. J. Quant. Chem*. **2000**, *77*, 895; the CASTEP program code was used. [Accelrys, Inc., San Diego, CA.]

[20] K. Ogasawara, T. Iwata, Y. Koyama, T. Ishii, I. Tanaka, H. Adachi *Phys. Rev. B* **2001**, *64*, 115413.

[21] Theoretical calculations of Mn-K EXAFS were done using the Wien2k code. [Blaha, P et al. WIEN2k, *An Augmented Plane Wave + Local Orbitals Program for Calculating Crystal Properties* (Karlheinz Schwarz, Techn. Universität Wien, Austria), **2001**. ISBN 3-9501031-1-2.]

[22] S. Sonoda, Y. Yamamoto, T. Sasaki, K. Suga, K. Kindo, H. Hori, *Solid State Comm*. **2005**, *133*, 177.

[23] T. Graf, M. Gjukic, M. S. Brandt, M. Stutzmann, O. Ambacher, *Appl. Phys. Lett*. **2002**, *81*, 5159. T. Graf, M. Gjukic, M. Hermann, M. S. Brandt, M. Stutzmann, L. Görgens, J. B. Philipp, O. Ambacher, *J. Appl. Phys*. **2003**, *93*, 9697.

[24] A. Wolos, M. Palczewska, M. Zajac, J. Gosk, M. Kaminska, A. Twardowski, M. Bockowski, I. Grzegory, S. Porowski, *Phys. Rev. B* **2004**, *69*, 115210.

[25] S. Limpijumnong, C. G. Van de Walle, *Phys. Rev. B* **2004**, *69*, 035207.

[26] D. C. Look, D. C. Reynolds, J. W. Hemsky, J. R. Sizelove, R. L. Jones, R. J. Molnar, *Phys. Rev. Lett*. **1997**, *79*, 2273.

[27] K. Shimomura, R. Kadono, K. Ohishi, M. Mizuta, M. Saito, K. H. Chow, B. Hitti, R. L. Lichti, *Phys. Rev. Lett*. **2004**, *92*, 135505.

[28] J. Neugebauer, C. G. Van de Walle, *Phys. Rev. Lett*. **1995**, *75*, 4452.

[29] K. H. Baik, *Appl. Phys. Lett*. **2003**, *83*, 5458.

[30] J. M. D. Coey, M. Venkatesan, C. B. Fitzgerald, *Nature Mater*. **2005**, *4*, 173.